\documentclass[12pt]{article}
\usepackage{graphicx} % Required for inserting images

\usepackage[small]{titlesec} % чуть уменьшаем размер заголовков глав и разделов
\usepackage[utf8]{inputenc} 
\usepackage[T1]{fontenc}
\usepackage[english]{babel}
\usepackage[left=2cm,right=2cm,top=2cm,bottom=2cm,bindingoffset=0cm]{geometry}

\graphicspath{{figures/}}

\usepackage{setspace}
\usepackage{subcaption}
\usepackage{youngtab}
\usepackage{epigraph}
\usepackage{upgreek}

\usepackage[labelfont=bf, margin=20pt,font={small}]{caption}

\usepackage{amsmath}
\usepackage{esint}
\usepackage{bm}
\usepackage{
    tikz,
	amsmath,
	amssymb,
	amsfonts,
	amsthm,
	amscd,
	comment,
	epsfig,
	color,
	setspace,
	bbold,
	verbatim,
	slashed
}
\usepackage{upgreek}
\usepackage{tikz}
\usetikzlibrary{arrows}
\usepackage{mathrsfs}
\usepackage{multirow}
\usepackage{simpler-wick}
\usepackage{cite}

\newcommand\beq{\begin{equation}}
\newcommand\eeq{\end{equation}}

\tikzset{cross/.style={cross out, draw=black, minimum size=2*(#1-\pgflinewidth), inner sep=0pt, outer sep=0pt},
	cross/.default={5pt}}
\usepackage{mathtools}
\usepackage{graphicx}
\usepackage{authblk}
\usepackage{xcolor}
\usepackage{bbold}
\usepackage[makeroom]{cancel}
\usepackage[hidelinks]{hyperref}

	\definecolor{linkcolor}{rgb}{0,0,1} % цвет ссылок
	\definecolor{urlcolor}{rgb}{0,0,1} % цвет гиперссылок
 
\hypersetup{pdfstartview=FitH, linkcolor=linkcolor,urlcolor=urlcolor,citecolor=urlcolor, colorlinks=true}

\author[1,2]{E.~T.~Akhmedov\thanks{\href{mailto:akhmedov@itep.ru}{akhmedov@itep.ru}}}
\author[1]{V.~I.~Lapushkin\thanks{\href{mailto:volapushkin@gmail.com}{volapushkin@gmail.com}}}
\author[1,3]{D.~I.~Sadekov\thanks{\href{mailto:sadekov.di@phystech.edu}{sadekov.di@phystech.edu}}}
\affil[1]{\itshape Institutskii per, 9, Moscow Institute of Physics and Technology, 141700, Dolgoprudny, Russia}
\affil[2]{\itshape Academician Kurchatov Square, 1, NRC ''Kurchatov Institute'', 123182, Moscow, Russia}
\affil[3]{\itshape P. N. Lebedev Physical Institute, Moscow 119991, Russia}
\title{\textcolor{black}{Light fields in various patches of de Sitter space-time}}

\begin{document}

\date{}
\maketitle

\begin{abstract}
    We start with the consideration of the loop effects for light fields with non-zero mass in the expanding Poincar\'e patch of de Sitter space-time. We derive the Dyson-Schwinger equation, which sums up the leading infrared (growing with time) loop corrections in certain limit for small initial perturbations above the Bunch-Davies state. The solution of this equation shows the destiny of the initial state at the future infinity. Then we discuss the case of the contracting Poincar\'e patch and global de Sitter space-time and briefly the case of different initial conditions in the expanding Poincar\'e patch. 
\end{abstract}

\newpage

\tableofcontents

\newpage
\section{Introduction}\label{Sec:Intro}

The problem of loop corrections in different expanding space-times has gained significant attention over the past two decades \cite{Starobinsky:1994bd, Tsamis:2005hd, Marolf:2010zp, Marolf:2010nz, Higuchi:2010xt, Moreau:2018lmz, Guilleux:2015pma, Hollands:2010pr, Gorbenko:2019rza, Akhmedov:2009vs, Krotov:2010ma, Akhmedov:2017ooy, Akhmedov:2019cfd}. First, quantum fluctuations in the very early Universe are believed to have determined its present large-scale structure \cite{Starobinsky:1980te,Starobinsky:1982ee,Linde:1981mu,Linde:1983gd,Guth:1980zm,Guth:1982ec,Albrecht:1982wi}. Second, resolving the cosmological constant problem may be connected to the large infrared (IR) contributions from loop corrections, as these corrections might lead to a secular screening of the cosmological constant \cite{Polyakov:2007mm, Akhmedov:2008pu, Polyakov:2009nq, Akhmedov:2009be, Polyakov:2012uc, Akhmedov:2013vka}. Third, understanding the backreaction of quantum matter on strong gravitational fields is incomplete without accounting for loop corrections (see e.g. for the reviews \cite{Akhmedov:2013vka}, \cite{Akhmedov:2021rhq}). Overall, it is of great interest to reduce our lack of understanding the role of quantum effects in interacting quantum field theory in various rapidly expanding backgrounds. These questions remain unresolved even in the case of de Sitter (dS) space-time, which possesses the highest degree of symmetry. The challenges arise due to the vacuum instability of dS \cite{Polyakov:2012uc}, the singular behavior of the light fields \cite{Tsamis:2005hd}, the breaking of global dS isometries at loop level \cite{Polyakov:2012uc}, \cite{Akhmedov:2013vka}.

In this paper, we continue investigating these questions, following the approach of \cite{Akhmedov:2009vs,Akhmedov:2013vka,Akhmedov:2017ooy,Akhmedov:2019cfd}. We mainly focus on the case of the massive real scalar fields from so called complementary series ($m<\frac{D-1}{2}$ in units of the Hubble constant) in expanding Poincar\'e patch (EPP), except for some discussion in Section \ref{Sec:Complementary_CPP} and Appendix. We discuss the differences in the situations in the EPP, contracting Poincar\'e patch (CPP) and in global dS. In order to set up the notations we first introduce the standard procedure of the quantization in dS and the Schwinger-Keldysh diagrammatic technique, which we will use to calculate the quantities we are interested in.

\subsection{Set up of the quantization in de Sitter space-time}\label{subSec:Quantization}

We consider the massive real scalar self-interacting theory:
\beq\label{eq:action}
    S[\phi] = \int d^{D}x\sqrt{|g|}\left[ \frac{1}{2}g^{\mu\nu}\partial_{\mu}\phi\partial_{\nu}\phi - \frac{1}{2}m^2\phi^2 - \frac{\lambda}{6}\phi^3\right],
\end{equation}
in the expanding Poincar\'e patch (EPP) of dS space-time:
\beq\label{eq:metric}
    ds^2 = \frac{1}{\eta^2}\left(d\eta^2 - d\bm{x}^2 \right),
\eeq
where we take the Hubble parameter to be unit $H=1$, and $\eta = e^{-t}$ is the conformal time in the EPP. It is ranging from $+\infty$ at the past infinity, $t\to - \infty$, to $0$ at the future one, $t\to +\infty$; $D=d+1$ is the space-time dimension. The cubic potential is taken to simplify equations below. The phenomena that we observe below have nothing to do with the instability of the cubic potential and are present also in quartic theory (see e.g. \cite{Akhmedov:2013vka}, \cite{Akhmedov:2013xka}). 

The free quantum scalar field in EPP can be expanded as:
\beq \label{eq:field_decomposition}
    \begin{aligned}
        \widehat{\phi}(\eta,\bm{x})=\int\dfrac{d^{D-1}\bm{p}}{(2\pi)^{D-1}}&\bigg[
\widehat{a}_{\bm{p}}f_{\bm{p}}(\eta)e^{i\bm{px}}+
\widehat{a}^{\dagger}_{\bm{p}}f^{*}_{\bm{p}}(\eta)e^{-i\bm{px}}
\bigg], \quad 
\left[\widehat{a}_{\bm{p}},\widehat{a}_{\bm{q}}^\dagger 
\right] = (2\pi)^{D-1}\delta(\bm{p}-\bm{q}),
\\
&{\rm where} \quad f_{\bm{p}}(\eta) = \eta^{\frac{D-1}{2}}h_{\nu}\left(p\eta\right),\;p\equiv |\bm{p}|.
    \end{aligned}
\eeq
The field operator obeys the canonical commutation relations with its conjugate momentum. To achieve the latter the modes $f_{\bm{p}}(\eta)e^{i\bm{px}}$ should obey the Klein-Gordon equation in the inflationary metric (\ref{eq:metric}). Then $h_{\nu}(p\eta)$ can e.g. be expressed in terms of the Hankel function of the first kind as follows:
\beq \label{eq:harmonics}
\begin{aligned}
    h_{\nu}\left(p\eta\right) &= \frac{\sqrt{\pi}}{2}H_{\nu}^{(1)}\left(p\eta\right),\; \nu = \sqrt{\frac{(D-1)^2}{4}-m^2}.
\end{aligned}
\eeq
These are the so called Bunch-Davies modes \cite{bunch1978quantum}. Other options are discussed at the end of the paper.

For the future use we show here the asymptotic form of these functions:
\beq\label{eq:asymptotics}
\begin{aligned}
    &h_{\nu}(p\eta) \simeq \frac{1}{\sqrt{2 p\eta}} e^{ip\eta}e^{-i\frac{\pi}{2}\left(\nu+\frac{1}{2}\right)}, \quad {\rm as} \quad p\eta \gg |\nu| ;
    \\
    &h_{\nu}(p\eta) \simeq iA_{-}\frac{1}{(p\eta)^{\nu}} + iA_{+}(p\eta)^{\nu} + iB(p\eta)^{-\nu+2} +\ldots, \quad {\rm as} \quad p\eta \ll |\nu|;
    \\
    &{\rm where} \quad A_{-} = -\frac{\Gamma(\nu)}{2\sqrt{\pi}}2^{\nu}, \; A_{+} = \frac{\sqrt{\pi}e^{-i\pi\nu}}{2^{\nu+1}\Gamma(\nu+1)\sin(\pi\nu)}, \; B = \frac{2^\nu}{8\sqrt{\pi}}\frac{\Gamma(\nu)}{2-\nu}.
\end{aligned}
\eeq
Note that the term $iB(p\eta)^{-\nu+2}$ and higher terms in the expansion on the right hand side in the second line are needed for the calculations below with small enough masses (such that $\nu$ is real and $\nu > 1$). Moreover, to simplify calculations below we assume that $\nu<\frac{D-1}{6}$ and $D\leq 4$. These restrictions will be explained below.

In dS space-time one distinguishes massive fields, $m>(D-1)/2$, corresponding to the pure imaginary $\nu$. They are from the so called principal series of the representations of the dS isometry group. In dS quantum field theory there are also light fields, $m<(D-1)/2$, corresponding to real $\nu$. They are from the so called complementary series. In this paper we will be mostly considering light fields, giving only a brief comment on the principal series in the Appendix.

\subsection{Schwinger-Keldysh formalism and the goal of the paper}\label{subSec:Keldysh_technique}
Below we perform calculations in the Schwinger-Keldysh diagrammatic technique after the Keldysh rotation (see e.g. \cite{kamenev2023field} for the review):
\beq\label{eq:Keldysh_rotation}
    \phi_{cl} = \frac{\phi_{+} + \phi_{-}}{2}, \; \phi_{q} = \phi_{+} - \phi_{-}.
\eeq
Then the interaction term in the theory under consideration takes the form
\beq\label{eq:Interaction_term}
    S_{\text{int}}[\phi_{cl},\phi_q] = S_{\text{int}}[\phi_{+}] - S_{\text{int}}[\phi_{-}] = \int d^{D}x\sqrt{|g|}\left[\frac{\lambda}{24}\phi_{q}^3 + \frac{\lambda}{2}\phi_{cl}^2\phi_{q} \right].
\eeq
Below in this technique we perform calculations with the Bunch-Davies state taken as the initial one at the light like initial Cauchy surface -- the boundary of the EPP. The state is specified by the condition $\widehat{a}_{\bm{p}}| \text{BD}\rangle = 0$ for the above defined modes and annihilation operators. 

In the technique in question the scalar field is characterized by the following three propagators:
\beq\label{eq:propagators_list}
    \begin{aligned}
        &iG^K_0(\eta_1,\bm{x}_1|\eta_2,\bm{x}_2) = \wick{\c\phi_{cl}(\eta_1,\bm{x}_1)\c\phi_{cl}(\eta_2,\bm{x}_2)} = \frac{1}{2}\bigg\langle \text{BD}\bigg| \bigg\{\widehat{\phi}(\eta_1,\bm{x}_1),\widehat{\phi}(\eta_2,\bm{x}_2) \bigg\} \bigg|\text{BD}\bigg\rangle,
        \\
        &iG^A_0(\eta_1,\bm{x}_1|\eta_2,\bm{x}_2) = \wick{\c\phi_{cl}(\eta_1,\bm{x}_1)\c\phi_{q}(\eta_2,\bm{x}_2)} = \theta(\eta_1-\eta_2)\rho(\eta_1,\bm{x}_1|\eta_2,\bm{x}_2),
        \\
        &iG^R_0(\eta_1,\bm{x}_1|t\eta_2,\bm{x}_2) = \wick{\c\phi_{q}(\eta_1,\bm{x}_1)\c\phi_{cl}(\eta_2,\bm{x}_2)} = -\theta(\eta_2-\eta_1)\rho(\eta_1,\bm{x}_1|\eta_2,\bm{x}_2), 
        \\
        &{\rm where} \quad \rho(\eta_1,\bm{x}_1|\eta_2,\bm{x}_2) =  \bigg[\widehat{\phi}(\eta_1,\bm{x}_1),\widehat{\phi}(\eta_2,\bm{x}_2) \bigg].
    \end{aligned}
\eeq
Here the subscript zero stands for the tree-level value of these two-point functions. The Bunch-Davies state is spatially homogeneous. Hence, it is convenient to perform the spatial Fourier transformation of the propagators. After the transformation we obtain that:
\beq\label{eq:propagators}
    iG^K_0(\bm{p}|\eta_1,\eta_2) = \text{Re}\left\{f_{\bm{p}}(\eta_1)f_{\bm{p}}^{*}(\eta_2)\right\}, \;\; 
    \rho(\bm{p}|\eta_1,\eta_2) = -2\text{Im}\left\{f_{\bm{p}}(\eta_1)f_{\bm{p}}^{*}(\eta_2)\right\}.
\eeq
The retarded and advanced propagators, $G_0^{R,A}$, are sensitive to the spectrum of the theory, but are not sensitive to the state of it (at the tree-level). Meanwhile, the Keldysh propagator, $G^K_0$, is sensitive to the state of the theory. 

Our goal is to trace the destiny of the state of the theory at the future infinity. That is the reason why we are interested in the behavior of the loop-corrected Keldysh propagator in the limit of large external times $\eta = \sqrt{\eta_1\eta_2} \to 0$, while $\eta_1/\eta_2 =$ const, in the units of the cosmological constant. In the momentum space (for each mode separately) this limit corresponds to $p\eta \ll |\nu|$. In this limit the loop corrected Keldysh propagator acquires the form \cite{Akhmedov:2013vka}:
\beq\label{eq:Keldysh_at_large_times}
    G^K(\bm{p}|\eta_1,\eta_2) \simeq \left(\frac{1}{2}+n_{p}(\eta)\right)f_{p}(\eta_1)f_{p}^*(\eta_2) + \varkappa_{p}(\eta) \, f_{p}(\eta_1)f_{p}(\eta_2) + \text{c.c},
\eeq
where in general $n_{p}(\eta)$ and $\varkappa_{p}(\eta)$ (perhaps multiplied by the mode functions as in (\ref{eq:Keldysh_at_large_times}) at the coincident points) are the level populations and the anomalous quantum averages, correspondingly. 

In the future infinity limit under consideration, $p\eta \to 0$, for the principal series ($m>\frac{D-1}{2}$) in the EPP the leading contributions to such products of modes as $f_{p}(\eta_1)f_{p}^*(\eta_2)$ and $f_{p}(\eta_1)f_{p}(\eta_2)$ are different functions of $\eta$'s, as follows from (\ref{eq:asymptotics}). Then, in such a case one can separate IR loop contributions to $n_{p}(\eta)$ and $\varkappa_{p}(\eta)$ from each other. However, for the complementary series in the EPP the leading contributions in the IR limit are the same, $f_{p}(\eta_1)f_{p}^*(\eta_2) \simeq -f_{p}(\eta_1)f_{p}(\eta_2)$. As a result, one obtains that
\beq\label{eq:Keldysh_large_times_complementary_EPP}
    G^K(\bm{p}|\eta_1,\eta_2) \simeq \frac{A_-^2 \eta^{D-1}}{(p\eta)^{2\nu}} N_{p}(\eta),
\eeq
where $N_{p}(\eta) = 1+2n_{p}(\eta)-2\text{Re}\left\{ \varkappa_{p}(\eta)\right\}$ and we have used asymptotic form of the modes (\ref{eq:asymptotics}). Moreover, in this case the leading first loop contributions to $n_{p}(\eta)$ and $\varkappa_{p}(\eta)$ cancel each other within $N_p(\eta)$, as is shown in \cite{Akhmedov:2017ooy}. We will show that below again.

As demonstrated in several works \cite{Krotov:2010ma,Akhmedov:2009vs, Akhmedov:2013vka,Akhmedov:2013xka,Akhmedov:2017ooy,Akhmedov:2019cfd}, in the future infinity limit, $p\eta \ll |\nu|$, of the EPP the Keldysh propagator at one-loop order (in $\phi^3$ theory) receives secularly growing contributions of the form $\sim \lambda^2\log(p\eta/|\nu|)$. They become comparable in magnitude to the tree-level contribution (\ref{eq:propagators_list}) for a long enough time of evolution of the system. This is true both for the heavy fields from the principal series and for the light ones from the complementary series (see, however, the discussion in the Appendix below). At the same time, one loop corrections to the retarded and advanced propagators are not growing as $p\eta \to 0$ and are just suppressed by $\lambda^2$. These large IR secular contributions drastically correct the physical effects seen at the tree-level, making it necessary to resum at least the leading multi-loop corrections of the form $\sim \left(\lambda^2\log(p\eta/|\nu|)\right)^n$ (assuming that $\lambda^2$ is small, while $\log(p\eta/|\nu|)$ is large). 

In this paper, we build upon and refine previous works on loop calculations for light fields in dS in the limit that was described above. Specifically, in the Sections \ref{subSec:One_loop_Complementary_EPP}, \ref{subSec:Multy-loop} and \ref{subSec:Vertices} we compute the leading secularly growing contributions at one loop level and demonstrate that multi-loop ``bubble inside bubble'' diagrams, as well as corrections to multi-point correlation functions, are suppressed. Based on these observations in the Section \ref{subSec:Resummation_complementary_EPP} we write down the Dyson-Schwinger equation, which resums the leading corrections from all loops in the limit under consideration in the EPP. These observations have already been made in \cite{Akhmedov:2019cfd}, but no correct resummation of the leading contributions was provided.  

In Section \ref{Sec:Complementary_CPP} we perform similar calculations for the light fields in the contracting Poincar\'e patch (CPP). The latter are having such a form as $\lambda^2\log\left(\frac{\eta}{\eta_0}\right)$. Here $\eta_0$ is the position of the initial Cauchy surface in the CPP. 

The case of CPP is discussed here for three reasons. First, the consideration of CPP is quite an important question even from the observational point of view and not only from the academic perspective. In fact, the same type of secular divergence as we just mentioned is present in the global dS space-time. That can be understood e.g. from the observation that the global dS is the union of EPP and CPP. At the same time we do not really know what was the initial state of our Universe at the start of the inflation. We do not know neither the initial Fock space state nor the geometry of the initial Cauchy surface. The state definitely was not the Bunch-Davies one at the light-like initial surface of the EPP. Otherwise even galaxies will not exist because the situation would be absolutely symmetric. The natural question is, once the start of the inflation was from a non symmetric state, what was achieved the first -- the equilibration or the strong backreaction on the background dS geometry \cite{Akhmedov:2021rhq}?

Second, in CPP we encounter a long-memory effect in dS \cite{Krotov:2010ma} (see also \cite{Akhmedov:2012dn,Akhmedov:2013vka}), manifesting itself as the secular divergence. To cut this divergence one needs to introduce into the loop corrections the dependence on the position of the initial Cauchy surface, $\eta_0$, thus, breaking the symmetry of the tree-level theory under the dS isometry group. This phenomenon was previously treated in works \cite{Akhmedov:2012dn,Akhmedov:2013vka} for the principal series in CPP and global dS within the kinetic approximation. There it was shown that whether the symmetry under the dS isometry is restored in the exact correlation functions or not depends on the initial conditions. 

Third reason to consider the situation in CPP is as follows. Unlike the case of the resummation in the EPP, the corresponding resumming equations in CPP are non-linear. The reason for that is as follows: due to the secular divergence different type of diagrams is providing the leading corrections. This fact making the resummation for light fields in CPP and global dS more challenging, particularly because the kinetic approximation is not applicable in this context. We will address this issue in future work. 

Additionally, in Appendix \ref{App:Principal_EPP}, we revisit the resummation of leading logarithms in EPP for the principal series. We argue that for heavy fields, loop corrections exhibit an additional suppression of the form $e^{-\pi\nu}$, on top of the $\lambda^2$, thus, behaving as $\sim e^{-\pi\nu}\lambda^2\log\left(p\eta/|\nu|\right)$. Finally, we show that for sufficiently large masses, $\nu \gtrsim \frac{2}{\pi}\log\left(\frac{1}{\lambda}\right)$, the level populations and the anomalous quantum averages attain small values at late times, resulting in only a slight perturbation to the initial Bunch-Davies state.

%\section{Complementary series in the expanding Poincare patch}\label{Sec:Complementary_EPP}
\section{One-loop corrections to the propagators}\label{subSec:One_loop_Complementary_EPP}

\begin{figure}[t]
    \centering
    \def\svgwidth{\textwidth}
    \includegraphics[width=\linewidth]{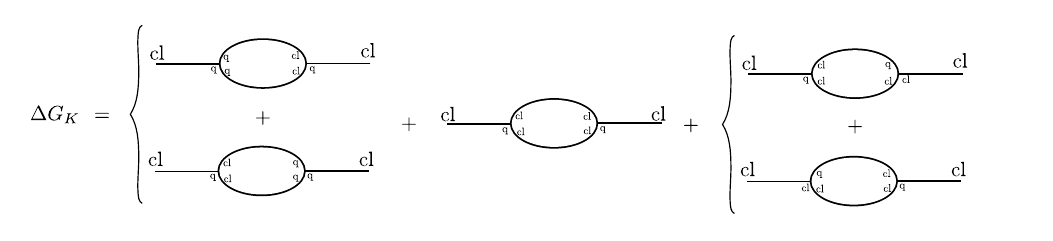}
    \caption{Bubble diagram.}
    \label{Fig:BubbleDiagram}
    \hfill
\end{figure}

In this section we single out the leading contribution from the one-loop correction to the Keldysh propagator in the limit $p\sqrt{\eta_1\eta_2} \equiv p\eta \to 0$ for the Bunch-Davies initial state. Having the tree-level vertices from (\ref{eq:Interaction_term}) and the propagators from (\ref{eq:propagators_list}) we can write all the relevant one-loop diagrams, which are shown on the Fig.\ref{Fig:BubbleDiagram}.
Keeping in mind that we are interested in secular effects in the limit, when $(t_1 + t_2)/2 \gg |t_1 - t_2|$, where $\eta_{1,2} = e^{-t_{1,2}}$, in the arguments of the corrected propagator \cite{Akhmedov:2019cfd}, we can neglect the difference between the external time coordinates. Then the leading one-loop correction in the momentum space is contained in the following expression:
\beq\label{eq:1-loop_correction_general_formula_EPP}
\begin{aligned}
    \Delta_{1\text{-loop}}G^{K}(\bm{p}|\eta,\eta) =  \lambda^2\int\frac{d^{D-1}\bm{k}}{(2\pi)^{D-1}}\frac{d\eta_{3}}{\eta_{3}^{D}}\frac{d\eta_{4}}{\eta_{4}^{D}}
    \bigg[-\frac{1}{2}\theta(\eta_3-\eta)\theta(\eta_4-\eta)\times&
    \\
    \times \rho\big(\bm{p}\big|\eta,\eta_3\big)\left(G^{K^2}_{0}\big(\bm{k}-\bm{p}\big|\eta_3,\eta_4\big) - \frac{1}{4}\rho^2\big(\bm{k}-\bm{p}\big|\eta_3,\eta_4\big)\right)\rho\big(\bm{p}\big|\eta_4,\eta\big) + &
    \\
    + \theta(\eta_4-\eta_3)\theta(\eta_3-\eta)\rho\big(\bm{p}\big|\eta,\eta_3\big)\rho\big(\bm{k}-\bm{p}\big|\eta_3,\eta_4\big) G^K_{0}\big(\bm{k}-\bm{p}\big|\eta_3,\eta_4\big)G^K_{0}\big(\bm{p}\big|\eta_4,\eta\big)
    + &
    \\
    +\theta(\eta_3-\eta_4)\theta(\eta_4-\eta)G^K_{0}\big(\bm{p}\big|\eta,\eta_3\big)\rho\big(\bm{k}-\bm{p}\big|\eta_3,\eta_4\big) G^K_{0}\big(\bm{k}-\bm{p}\big|\eta_3,\eta_4\big)\rho\big(\bm{p}\big|\eta_4,\eta\big)
    \bigg]&.
\end{aligned}
\eeq
First, the analysis of the expression (\ref{eq:1-loop_correction_general_formula_EPP}) shows that the dominant contribution in the limit under consideration comes from the region of the momentum integrals, where the external momentum, $\bm{p}$, can be neglected in comparison with the internal one, $\bm{k}$ \cite{Krotov:2010ma}. Furthermore, to single out the leading correction one can assume that the internal conformal times, $\eta_3, \eta_4$, are such that the IR asymptotic form (\ref{eq:asymptotics}) for the modes can be used in the external propagators \cite{Krotov:2010ma}. It must be also noted that in doing the latter we must assume the condition that $\nu < \frac{D-1}{4}$ for the IR convergence of these expressions with $\bm{p}$ taken to zero\footnote{Actually, an accurate calculation shows that there is nothing much different happens when $\nu>\frac{D-1}{4}$, but the coefficients in the resulting expressions are slightly changed.}. 

Second, the terms in the last two lines of (\ref{eq:1-loop_correction_general_formula_EPP}) are actually equal after the exchange of the integration variables $\eta_3\leftrightarrow \eta_4$. Let us denote the sum of these two terms as $\Delta_{1}G^K(\bm{p}|\eta,\eta)$ and use the expressions for the propagators via the modes to write the expression in the limit discussed above as:
\beq\label{eq:dominant_contr_1-loop_EPP_1}
\begin{aligned}
     \Delta_{1}G^K(\bm{p}|\eta,\eta) \simeq -\frac{2A_{-}^2}{(p\eta)^{2\nu}}\eta^{D-1}\lambda^2\int_{1}^{\infty}\frac{d\tau_4}{\tau_4}\int_{\tau_4}^{\infty}\frac{d\tau_3}{\tau_3}(\tau_3\tau_4)^{\frac{D-1}{2}}\text{Im}\big\{h_{\nu}(p\eta\tau_3)\big\}\text{Re}\big\{h_{\nu}(p\eta\tau_4)\big\}\times
     \\
     \times \int\frac{d^{D-1}\bm{q}}{(2\pi)^{D-1}}\text{Im}\bigg\{h_{\nu}^{2}\left(q\tau_3\right)h_{\nu}^{*2}\left(q\tau_4\right)\bigg\},
\end{aligned}
\eeq
where we have made the following change of the integration variables $\tau_3 = \frac{\eta_3}{\eta},\;\tau_4 = \frac{\eta_4}{\eta},\;\bm{q} = \bm{k}\eta$. Let us now make the further change of variables $u=\sqrt{\tau_3\tau_4}, \;v=\sqrt{\frac{\tau_3}{\tau_4}},\;\bm{\ell}=\bm{q}u$ and use the IR asymptotic form (\ref{eq:asymptotics}) for $h_{\nu}(p\eta\tau_3)$ and $h_{\nu}(p\eta\tau_4)$, to obtain that
\beq\label{eq:dominant_contr_1-loop_EPP}
\begin{aligned}
    \Delta_{1}G^K(\bm{p}|\eta,\eta) \simeq \frac{8A_{-}^3\text{Im}\{A_{+}\}}{(p\eta)^{2\nu}}\eta^{D-1}\int_{1}^{\frac{1}{\sqrt{p\eta}}}\frac{dv}{v^{2\nu+1}}\int_{v}^{\frac{1}{p\eta v}}\frac{du}{u}\int\frac{d^{D-1}\bm{\ell}}{(2\pi)^{D-1}}\text{Im}\left\{h_{\nu}^2\big(\ell v\big)h_{\nu}^{*2}\left(\frac{\ell}{v}\right)\right\} \simeq
    \\
    \simeq \frac{A_{-}^2}{(p\eta)^{2\nu}}\eta^{D-1} \left[A\cdot\lambda^2\log\left(\frac{p\eta}{|\nu|}\right) + B\cdot\lambda^2\right],
\end{aligned}
\eeq
where
\beq\label{defFv}
    A \equiv -8A_{-}\text{Im}\{A_+\}\int_{1}^{\infty}\frac{dv}{v^{2\nu+1}}\text{Im}\big\{F(v)\big\},\;\;
    F(v) \equiv \int\frac{d^{D-1}\bm{\ell}}{(2\pi)^{D-1}}\text{Im}\left\{h_{\nu}^2\big(\ell v\big)h_{\nu}^{*2}\left(\frac{\ell}{v}\right)\right\},
\eeq
and $B$ is some irrelevant constant for the further discussion. To obtain the second line of (\ref{eq:dominant_contr_1-loop_EPP}) we set the upper limit of the integration over $v$ to infinity due to the rapid convergence of these integrals on much shorter intervals. One should keep in mind that we mainly restrict ourselves to the dimensions $D\leq 4$. 

We see that for sufficiently large times (or low conformal times in EPP) the logarithm in (\ref{eq:dominant_contr_1-loop_EPP}) becomes large and the correction can dominate even over the tree-level Keldysh propagator, while the other contributions are suppressed as $\mathcal{O}\left(\lambda^2\right)$ in this limit. In the same way it can be shown that other terms in (\ref{eq:1-loop_correction_general_formula_EPP}) give no additional large or growing with time contributions. Hence, in the one-loop order the dominant correction to the Keldysh propagator (\ref{eq:1-loop_correction_general_formula_EPP}) is given by
\beq\label{eq:correction_1-loop_EPP}
    \Delta_{1-\text{loop}}G^{K}(\bm{p}|\eta,\eta) \simeq G^{K}_{0}(\bm{p}|\eta,\eta)\cdot A\lambda^2\log\left(\frac{p\eta}{|\nu|}\right), \quad {\rm as} \quad p\eta \to 0.
\eeq
This is the secular correction that we have been talking about in the Introduction. It should be stressed at the same time that retarded and advanced propagators do not receive growing with time contributions in the limit $\sqrt{\eta_1\eta_2} \to 0$, $\eta_1/\eta_2 = const$ \cite{Akhmedov:2013vka}. Thus, only the Keldysh propagator contains the secular contribution, which, hence, cannot be e.g. absorbed into the self-energy and/or mode frequency shift. 

\section{Multi-loop contributions to the two-point functions}\label{subSec:Multy-loop}

At late times, when $\lambda^2\log(p\eta) \sim 1$, the perturbation expansion breaks down and we need to sum up at least the leading secular corrections from all loops. In order to write the system of appropriate Dyson-Schwinger equations, which does the summation of the leading corrections, we have to estimate which type of diagrams should be taken into account. Obviously, there are ladder (bead or chaplet type) diagrams which will give powers of the form $\left[\lambda^2\log(p\eta)\right]^n$, $n\geq 2$. 

\begin{figure}[t]
    \centering
    \def\svgwidth{\textwidth}
    \includegraphics[width=16cm]{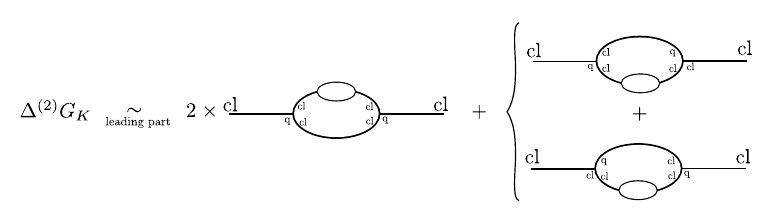}
    \caption{Bubble inside bubble.}
    \label{Fig:BubbleInsideBubble}
    \hfill
\end{figure}

In this section we show that the ``bubble inside bubble'' diagrams, such as those shown on Fig.\ref{Fig:BubbleInsideBubble}, contribute subleading corrections. In view of the fact that one-loop corrections to the retarded and advanced propagators do not contain any secular growth, one simple way to see this fact is to substitute the leading one loop correction (\ref{eq:correction_1-loop_EPP}) into the internal Keldysh lines of Fig.\ref{Fig:BubbleDiagram}.
Estimating the correction in the internal line of the diagrams as 
\beq\label{eq:estimation_internal_G^K}
\Delta G^K(\bm{q}|\eta_3,\eta_4) \sim \lambda^2\log(q\sqrt{\eta_3\eta_4})G^K(\bm{q}|\eta_3,\eta_4) = \lambda^2\log(\ell)G^K(\bm{q}|\eta_3,\eta_4),
\eeq
we get
\beq\label{eq:estim_BubbleInsideBubble}
    \begin{aligned}
        \Delta^{(2)} G^K(\bm{p}|\eta,\eta) \sim \frac{\lambda^4}{(p\eta)^{2\nu}}\int_{1}^{\frac{1}{\sqrt{p\eta}}}\frac{dv}{v^{2\nu+1}}\int_{v}^{\frac{1}{p\eta v}}\frac{du}{u}\int d^{D-1}\bm{\ell}\log(\ell)h_{\nu}^2(\ell v)h_{\nu}^{*2}\left(\frac{\ell}{v}\right) \sim&
        \\
         \sim \frac{\lambda^4 \, \eta^{D-1}}{(p\eta)^{2\nu}}\log(p\eta) \sim \lambda^4\log(p\eta)G^K_{0}(\bm{p}|\eta,\eta)&.
    \end{aligned}
\eeq
The resulting expression (\ref{eq:estim_BubbleInsideBubble}) is, thus, suppressed by the additional power of $\lambda^2$, which is not accompanied by an additional power of the logarithm. This very important remark was not taken into account in \cite{Akhmedov:2013vka, Akhmedov:2017ooy}. Then it was corrected in \cite{Akhmedov:2019cfd}, but without the revision of the Dyson-Schwinger equation and its solutions. In this work we fill in this loophole in the research. The point is that if the diagrams shown on the Fig.\ref{Fig:BubbleInsideBubble} are included in the leading approximation, then the resulting integral equation on the corrected Keldysh propagator is not linear and has redundant singular solutions \cite{Akhmedov:2017ooy}. Nevertheless, in \cite{Akhmedov:2013vka} it was shown that non-linear equations resumming diagrams appear in CPP and global dS. That was shown for the case of principal series, and we will demonstrate the same for the light fields from the complementary series below.

\section{Corrections to multi-point correlation functions}\label{subSec:Vertices}

\begin{figure}[t]
    \centering
    \def\svgwidth{\textwidth}
    \includegraphics[width=\textwidth]{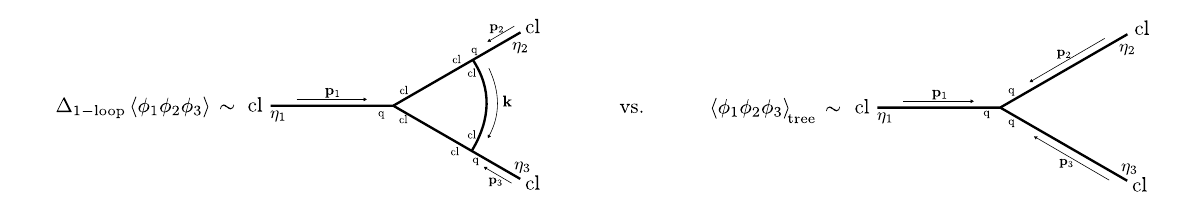}
    \caption{One loop correction to the three-point function to be compared with the tree-level value in the limit $p_{1,2,3}\eta_{1,2,3}\rightarrow 0$. The subscripts of the fields in the correlators symbolically mean the corresponding time coordinate and the diagrams are written in the momentum representation.}
    \label{Fig:Vertex_1-loop}
    \hfill
\end{figure}

The system of Dyson-Schwinger equations is imposed on propagators and multipoint correlation functions. Hence, to single out the leading corrections we should check for potentially dangerous (growing with time) loop contributions to higher-point correlators as well. At the one loop level the correction to the three-point correlation function consists of diagrams shown on Fig.\ref{Fig:Vertex_1-loop}.

The loop corrections from the Fig.\ref{Fig:Vertex_1-loop} have the following form modulo the overall coefficient:
\beq\label{eq:vertex_1-loop-general_formula}
\begin{aligned}
        \Delta_{1-\text{loop}}\left\langle\phi_1\phi_2\phi_3\right\rangle \sim 
        \\
        \sim \delta^{(D-1)}\left(\bm{p}_1+\bm{p}_2+\bm{p}_3\right)\cdot\lambda^{3}\cdot\int_{\eta_1}^{\infty}\frac{d\eta_4}{\eta_4^{D}}\int_{\eta_2}^{\infty}\frac{d\eta_5}{\eta_5^{D}}\int_{\eta_3}^{\infty}\frac{d\eta_6}{\eta_6^{D}}\rho\big(\bm{p}_1\big|\eta_1,\eta_4\big)\rho\big(\bm{p}_2\big|\eta_5,\eta_2\big)\rho\big(\bm{p}_3\big|\eta_6,\eta_3\big)\times
        \\
        \times \int\frac{d^{D-1}\bm{k}}{(2\pi)^{D-1}}G^K_{0}\big(\bm{k}-\bm{p}_2\big|\eta_4,\eta_5\big)G^K_{0}\big(\bm{k}\big|\eta_5,\eta_6\big)G^K_{0}\big(\bm{k}+\bm{p}_3\big|\eta_6,\eta_4\big).
\end{aligned}
\eeq
There are potentially large IR contributions from the loop momentum integral in (\ref{eq:vertex_1-loop-general_formula}) in the case $\nu>\frac{D-1}{6}$ in view of the asymptotic behavior (\ref{eq:asymptotics}). To make an estimation of this correction let us take the integral in one region where all propagators of this diagram can be expanded at small arguments: $k<\text{min}\left\{\frac{1}{\eta_5},\frac{1}{\eta_6},p_2,p_3\right\},\;\eta_4<\frac{1}{\text{max}\left\{p_1,p_2,p_3\right\}},\;\eta_5<\frac{1}{p_2},\;\eta_6<\frac{1}{p_3}$:
\beq\label{eq:vertex_1-loop_estimation}
    \begin{aligned}
        \Delta_{1-\text{loop}}\left\langle\phi_1\phi_2\phi_3\right\rangle \sim A_{-}^9\text{Im}^{3}\left\{A_{+}\right\}\delta^{(D-1)}\left(\bm{p}_1+\bm{p}_2+\bm{p}_3\right)\cdot\lambda^{3}\cdot\frac{ (\eta_1\eta_2\eta_3)^{D-1-\nu}}{(p\eta_3)^{\nu}(p\eta_2)^{\nu}(p\eta_3)^{\nu}}\times
        \\
        \times \frac{p_1^{\nu}\text{min}^{D-1-2\nu}\big\{p_2,p_3\big\}}{p_2^{\nu}p_1^{\nu}}\frac{1}{\left(\eta_1\text{max}\big\{p_1,p_2,p_3\big\}\right)^{\frac{D-1}{2}-\nu}}\frac{1}{\left(p_2\eta_2\right)^{\frac{D-1}{2}-\nu}}\frac{1}{\left(p_3\eta_3\right)^{\frac{D-1}{2}-\nu}}\;\;.
    \end{aligned}
\eeq
At the same time the typical contribution to the three-point function at tree-level (the second picture in Fig.\ref{Fig:Vertex_1-loop}) has the form
\beq
\begin{aligned}
    \left\langle\phi_1\phi_2\phi_3\right\rangle_{\text{tree}} \sim \delta^{(D-1)}\left(\bm{p}_1+\bm{p}_2+\bm{p}_3\right)\cdot\lambda\cdot\int\frac{d\eta_4}{\eta_4^{D}}G^{A}\big(\bm{p}_1\big|\eta_1,\eta_4\big)G^{R}\big(\bm{p}_2\big|\eta_4,\eta_3\big)G^{R}\big(\bm{p}_3\big|\eta_4,\eta_3\big)\sim
    \\
    \sim \delta^{(D-1)}\left(\bm{p}_1+\bm{p}_2+\bm{p}_3\right)\cdot\lambda\cdot \frac{(\eta_1\eta_2\eta_3)^{\frac{D-1}{2}}}{(p_1\eta_1)^{\nu}(p_2\eta_2)^{\nu}(p_3\eta_3)^{\nu}}\frac{(p_1p_2p_3)^{\nu}}{\left(\text{max}\big\{p_1,p_2,p_3\big\}\right)^{\frac{D-1}{2}+3\nu}}.
\end{aligned}
\eeq
Now let us introduce dimensionless variables $x_{1,2,3}$ and $\tau_{1,2,3}$ through $p_{1,2,3}=px_{1,2,3},\;\eta_{1,2,3}=\eta\tau_{1,2,3}$. As in the case of the Keldysh propagator we are interested in the secular effects at late times. Hence, we choose the external parameters to be of the order $x_{1,2,3}\sim 1$ and $\tau_{1,2,3}\sim 1$ and take the limit $p\eta\rightarrow 0$. Then it is evident that the one loop correction is just suppressed in this limit:
\beq\label{eq:vertex_1-loop_VS_tree}
    \frac{\Delta_{1-\text{loop}}\left\langle\phi_1\phi_2\phi_3\right\rangle}{ \left\langle\phi_1\phi_2\phi_3\right\rangle_{\text{tree}}}\sim \lambda^2\cdot\frac{\eta^{3(D-1-\nu)}}{(p\eta)^{3\nu}}\frac{p^{D-1-3\nu}}{(p\eta)^{3\frac{D-1}{2}-3\nu}}\bigg/\frac{\eta^{3\frac{D-1}{2}}}{(p\eta)^{3\nu}}\frac{p^{3\nu}}{p^{\frac{D-1}{2}+3\nu}}\sim \lambda^2,
\eeq
More accurate estimates of the other corrections to the three-point functions also show no additional growing with time contributions in the limit under consideration. It happens because in the multi-point correlation functions the ``volume'' factors $\eta^{\frac{D-1}{2}}$ in the modes (\ref{eq:field_decomposition}) and propagators (\ref{eq:propagators}) suppress the peculiar IR behavior of the light fields (\ref{eq:asymptotics}) when integrated over the internal conformal times of the diagrams. 

Thus, we do not need to write the Dyson-Schwinger equations for the multi-point correlation functions when we sum up the leading secularly growing corrections of the form (\ref{eq:dominant_contr_1-loop_EPP}), and can keep these correlation functions of the tree-level form in the equations for the propagators. 

However, let us stress that there is a strong IR correction at the one-loop order for the vertex function, which is the amputated diagram of the form shown on Fig.\ref{Fig:Vertex_1-loop}, i.e. with removed external propagators in the expression (\ref{eq:vertex_1-loop-general_formula}). In fact, in this case the corresponding estimates give:
\beq\label{eq:amputated_vertices_estimation}
\begin{aligned}
        \Delta_{1-\text{loop}}\left\langle\phi_1\phi_2\phi_3\right\rangle^{\text{Amp.}} \sim \delta^{(D-1)}\left(\bm{p}_1+\bm{p}_2+\bm{p}_3\right) \cdot\lambda^{3}\cdot(\eta_1\eta_2\eta_3)^{D-1-2\nu}\frac{\text{max}\big\{p_2,p_3\big\}}{p_2^{2\nu}p_3^{2\nu}},
         \\
        \left\langle\phi_1\phi_2\phi_3\right\rangle_{\text{tree}}^{\text{Amp.}}\sim \delta^{(D-1)}\left(\bm{p}_1+\bm{p}_2+\bm{p}_3\right)\cdot \lambda\cdot\delta(\eta_1-\eta_2)\delta(\eta_2-\eta_3)(\eta_1\eta_2)^{D}.
\end{aligned}
\eeq
Hence, in the limit $p\eta\rightarrow 0$ the ratio of these two contributions is as follows:
\beq\label{eq:relation_vertices_1-loopVStree}
    \frac{\Delta_{1-\text{loop}}\left\langle\phi_1\phi_2\phi_3\right\rangle^{\text{Amp.}}}{\left\langle\phi_1\phi_2\phi_3\right\rangle_{\text{tree}}^{\text{Amp.}}} \sim \lambda^2(p\eta)^{D-1-6\nu},
\eeq
which is singular if $\nu>\frac{D-1}{6}$. In \cite{Akhmedov:2017ooy} it is pointed out that such a behavior of the vertex functions is connected to the appearance of the discrete set of states from the complementary series in the K\"all\'en-Lehmann decomposition for small enough masses of the scalar field. Nevertheless, as we discussed above, these observations do not complicate the procedure of the resummation of the leading contributions of the form of powers of (\ref{eq:correction_1-loop_EPP}). 

\section{Resummation}\label{subSec:Resummation_complementary_EPP}

The observations that have been made in the previous sections allow us to state that at the leading approximation in the EPP in the limit $p\sqrt{\eta_1\eta_2} \rightarrow 0$, $\eta_1/\eta_2 = const$ we should sum up the bead or chaplet type diagrams constructed from the one depicted on Fig.\ref{Fig:BubbleDiagram}. Such a summation has to be done only for the Keldysh propagator, while retarded and andavced propagators, and vertexes can be taken of the tree-level form. (We take care only about the leading secular IR contributions and assume that all the constants in the theory are having their physical values, i.e. are UV renormalized.) 

The corresponding Dyson-Schwinger equation is
\beq\label{eq:DS-equation_Complementary_EPP}
    \begin{aligned}
    G^{K}(\bm{p}|\eta,\eta) \simeq G^{K}_{0}(\bm{p}|\eta,\eta)+ \lambda^2\int\frac{d^{D-1}\bm{k}}{(2\pi)^{D-1}}\frac{d\eta_{3}}{\eta_{3}^{D}}\frac{d\eta_{4}}{\eta_{4}^{D}}
    \bigg[-\frac{1}{2}\theta(\eta_3-\eta)\theta(\eta_4-\eta)\times&
    \\
    \times \rho\big(\bm{p}\big|\eta,\eta_3\big)\left(G^{K^2}_{0}\big(\bm{k}-\bm{p}\big|\eta_3,\eta_4\big) - \frac{1}{4}\rho^2\big(\bm{k}-\bm{p}\big|\eta_3,\eta_4\big)\right)\rho\big(\bm{p}\big|\eta_4,\eta\big) + &
    \\
    + \theta(\eta_4-\eta_3)\theta(\eta_3-\eta)\rho\big(\bm{p}\big|\eta,\eta_3\big)\rho\big(\bm{k}-\bm{p}\big|\eta_3,\eta_4\big) G^K_{0}\big(\bm{k}-\bm{p}\big|\eta_3,\eta_4\big)G^K\big(\bm{p}\big|\eta_4,\eta\big)
    + &
    \\
    +\theta(\eta_3-\eta_4)\theta(\eta_4-\eta)G^K_{0}\big(\bm{p}\big|\eta,\eta_3\big)\rho\big(\bm{k}-\bm{p}\big|\eta_3,\eta_4\big) G^K_{0}\big(\bm{k}-\bm{p}\big|\eta_3,\eta_4\big)\rho\big(\bm{p}\big|\eta_4,\eta\big)
    \bigg]&.
\end{aligned}
\eeq
On the RHS of this equation the exact form has only the Keldysh propagator and only in one place.

In order to solve this equation, we will use the following ansatz for the ``exact'' Keldysh propagator:
\beq\label{eq:ansatz_for_GK_1}
    G^K(\bm{p}|\eta_1,\eta_2) = A_{-}^2 \eta^{D-1}\frac{N\left(p\eta\right)}{(p\eta)^{2\nu}}, \; \eta = \sqrt{\eta_1\eta_2},
\eeq
which solves this equation under certain conditions in the limit that we consider, as we will see now. In general, we are interested in solving the equation (\ref{eq:DS-equation_Complementary_EPP}) for different initial conditions for the Keldysh propagator, which carries the information about the quantum state of the scalar field at the exit surface\footnote{One should keep in mind that loop corrections respect the dS isometry in the EPP for the BD state \cite{polyakov2012infraredinstabilitysitterspace}, \cite{Higuchi_2011}, \cite{hollands2010correlatorsfeynmandiagramsquantum}, \cite{Akhmedov:2022uug}. But it is more interesting to trace the destiny of perturbations (perhaps spatially homogeneous) of the BD state that violate the dS isometry at the initial moment.
} $p\eta \sim |\nu|$. 

In this paper we use the simplest choice. Namely, we assume some initial value for $N(p\eta)$ in (\ref{eq:ansatz_for_GK_1}), $N_0 \equiv N(P_0) = 1+2n(P_0)-2\text{Re}\left\{\varkappa(P_0)\right\}$, which is imposed for some initial physical momentum $P_0 = (p\eta)_0$. Such a choice preserves at least the following part of the dS isometry\footnote{It must be stressed here that  this is a rather specific choice of the initial conditions from the physical perspective (see e.g. \cite{Akhmedov:2024ice} for the related discussion). There are many more different ways to set up initial conditions, and they normally break the dS isometry. This point is discussed in a bit more details in the concluding section of this paper.}: $p\rightarrow \frac{1}{\alpha}p,\;\eta\rightarrow \alpha \eta$ for some constant $\alpha$. Therefore, the resulting equation for the evolution of the quantum state takes the form
\beq\label{eq:DS_for_N_Complementary_EPP}
    N(p\eta) \simeq N_0 + 4 \, A_{-} \, \text{Im}\{A_{+}\}\, N_{0}\,\lambda^2\int_{1}^{\frac{1}{\sqrt{p\eta}}}\frac{dv}{v^{2\nu+1}}\int_{v}^{\frac{1}{p\eta v}}\frac{du}{u}\Big[N(p\eta\sqrt{uv}) + N_{0}\Big]\text{Im}\big\{F(v)\big\} + \mathcal{O}(\lambda^2),
\eeq
where $A_\pm$ are defined in (\ref{eq:asymptotics}) and $F(v)$ is defined in (\ref{defFv}).
Let us stress again that the equation is valid in the approximations discussed in Section \ref{subSec:One_loop_Complementary_EPP}. 

The equation (\ref{eq:DS_for_N_Complementary_EPP}) in principle may have different solutions for different initial conditions. An obvious ansatz for the solution can be taken of the form $N(p\eta) = C(p\eta)^{\alpha}$. If $\alpha > 0$, then $N(p\eta\sqrt{uv})\ll N_0$ in eq. (\ref{eq:DS_for_N_Complementary_EPP}), as $p\eta \to 0$, and, hence, the ansatz is not appropriate, because it does not solve the equation under consideration. 

In contrast, if we assume that $\alpha < 0$, then $N(p\eta\sqrt{uv})\gg N_0$ and we obtain from (\ref{eq:DS_for_N_Complementary_EPP}) the equation defining $\alpha$:
\beq\label{eq:DS_for_alpha_1}
    \alpha \simeq -4\lambda^2 N_0 A_{-}\text{Im}\left\{A_{+}\right\}\int_{1}^{\infty}\frac{dv}{v^{2\nu+1}}\text{Im}\left\{F(v)\right\}v^{\frac{\alpha}{2}},
\eeq
where we set the upper limit of the integral over $v$ to infinity due to its rapid convergence on much smaller interval. In the leading order the solution for $\alpha$ is very close to zero:

\beq\label{eq:Solution_for_alpha_EPP}
    \alpha \simeq -4\lambda^2 N_0 A_{-}\text{Im}\left\{A_{+}\right\}\int_{1}^{\infty}\frac{dv}{v^{2\nu+1}}\text{Im}\left\{F(v)\right\} + \mathcal{O}(\lambda^4),
\eeq
providing that the initial value $N_0$, which can have both signs, should be such that $\alpha<0$. In this case the exact Keldysh propagator blows up in comparison with its tree-level value. In fact, the rate is: 
$$
\frac{G^K(\bm{p}|\eta, \eta)}{G^K_0(\bm{p}|\eta,\eta)} \sim (p\eta)^{-|\alpha|}.
$$ 
Nevertheless this power-law growth of the Keldysh propagator doesn't overcome the volume factor, $\eta^{D-1}$, in (\ref{eq:field_decomposition}), (\ref{eq:propagators}) or (\ref{eq:ansatz_for_GK_1}) and generally can be considered as an IR renormalization of the mass, making the situation somewhat similar to the one considered in \cite{Gautier:2013aoa,Guilleux:2016oqv,Serreau:2013psa} in large-$N$ models. (The difference in details of our answer with respect to the one in those papers can be attributed to the fact that we consider a different limit.) We will return to the Dyson-Schwinger equation (\ref{eq:DS-equation_Complementary_EPP}) to consider more generic initial states and types of self-interactions in our future works.

\section{A comment on the situation in the contracting Poincare patch}\label{Sec:Complementary_CPP}

In order to illustrate the situation where the dS isometry is inevitably broken at the loop order (even if at the tree-level it was respected) we examine here similar calculations in CPP for the complementary series. 

In CPP we consider the same initial Bunch-Davies state and the corresponding modes (\ref{eq:harmonics}) as in EPP, i.e. we also consider spatially homogeneous state, which of course in CPP, unlike the case of EPP, is unstable with respect to spatially inhomogeneous perturbations. However, the effects that we consider below have nothing to do with this instability and are present e.g. in the upper part of the global dS space-time.

The metric in CPP being expressed in conformal time is the same as in EPP, but the key difference is now the relation between conformal and cosmological time, $\eta=e^{t}$, and the arguments of theta-functions in (\ref{eq:propagators_list}) must be swapped between 1 and 2. Hence, the conformal time, $\eta$, is ranging from $\eta=0$ at past infinity, $t=-\infty$, to $\eta = +\infty$ in the future infinity, $t = +\infty$.

This choice of the mode functions and the state in CPP corresponds to the propagators that respect the proper Hadamard behavior and the dS isometry at the tree level. But this is not the only option for the initial state. In fact, it may be appropriate to consider in-modes in CPP, which are proportional to the Bessel function $h_{\nu}\sim Y_{\nu}(p\eta)$ (see e.g. \cite{Akhmedov:2013xka, Akhmedov:2013vka}) rather than Hankel ones. This Bessel function is some linear combination of the Hankel functions $H_{\nu}^{(1)}$ and $H_{\nu}^{(2)}$. (In the case of the principal series these modes behave as single oscillating exponents of the cosmological time at past infinity.) Then the corresponding (to the Fock space ground state) propagators also respect the dS isometry, but they do not have the proper Hadamard behavior. Furthermore, one should keep in mind that for other so called alpha-states, different from the BD one, which respect the dS isometry at tree-level, this symmetry is broken in the loops \cite{Akhmedov:2022uug} even in EPP. In a certain sense the latter fact is related to the non-Hadamard behavior of the corresponding propagators. 

In the case of CPP we encounter the secular divergence \cite{Krotov:2010ma}, \cite{Akhmedov:2019cfd} rather than the secular growth. In fact, one can straightforwardly see from (\ref{eq:1-loop_correction_general_formula_EPP}) that in EPP the $\eta_{3,4}$ integrals are bounded from below by the external time $\eta$ due to the causality of the Schwinger-Keldysh technique. Then the resulting expression for the correction in EPP (\ref{eq:correction_1-loop_EPP}) is singular in the limit $\eta \rightarrow 0$. 
Now, in the CPP the external conformal time, $\eta = \sqrt{\eta_1\eta_2}$, is the upper limit of the $\eta_{3,4}$ integrals rather than the lower one. Then the analogs of integrals in (\ref{eq:1-loop_correction_general_formula_EPP}) for CPP are divergent, because the lower bound in them is zero. To cut the divergence we must introduce the lower bound of $\eta_{3,4}$ integration by imposing some initial conformal time $\eta_0$. This is the position of an initial Cauchy surface that cannot be taken it to zero (or $t_0 = \log \eta_0$ to past infinity) because of the singularity. This introduction of the initial Cauchy surface breaks the dS isometry at one loop level. 

Then the reasonable question arises whether this symmetry is restored after the resummation of the loop corrections or not, i.e. can one remove the cut $\eta_0$ from the loop-corrected answer or not? 
In the loop calculations in CPP for the light fields there are two situations that should be considered separately. First situation is appears for the modes with momentum $p \sqrt{\eta_1\eta_2} = p\eta > |\nu|$. Then in the external legs we cannot use for the Hankel functions their approximate form for small arguments (\ref{eq:asymptotics}). Hence, one can distinguish between the quantum level populations $n_{p}$ and the anomalous quantum average $\varkappa_{p}$ inside the whole correction to the Keldysh propagator as in (\ref{eq:Keldysh_at_large_times}). To estimate the leading corrections to $n_{p}$ and $\varkappa_{p}$, as $p\eta_0 \to 0$,
one can use for the Hankel functions inside $\eta_{3,4}$ integrals their asymptotic form for small arguments (\ref{eq:asymptotics}) when $\eta_{3,4}\leq\frac{|\nu|}{p}$. Then the gained contributions to these quantities in the one-loop order are as follows:
\beq\label{eq:n_p_one-loop_CPP}
    \begin{aligned}
        n_p(\eta) \simeq \frac{\lambda^2}{(p\eta_0)^{2\nu}}\cdot\frac{A_{-}^2}{\nu}\int_{1}^{\infty}\frac{dv}{v^{2\nu+1}}\text{Re}\left\{F(v)\right\} -
        \\
        -2 \lambda^2\log\left(\frac{p\eta_0}{|\nu|}\right)\cdot A_{-}\int_{1}^{\infty}\frac{dv}{v}\text{Re}\left\{A_{+}v^{2\nu} + A_{+}^{*}\frac{1}{v^{2\nu}}F(v)\right\} +\mathcal{O}(\lambda^2),
    \end{aligned}
\eeq
\beq\label{eq:kappa_one-loop_CPP}
    \begin{aligned}
        \varkappa_p(\eta) \simeq \frac{\lambda^2}{(p\eta_0)^{2\nu}}\cdot\frac{A_{-}^2}{\nu}\int_{1}^{\infty}\frac{dv}{v^{2\nu+1}}F(v) -
        \\
        -2 \lambda^2\log\left(\frac{p\eta_0}{|\nu|}\right)\cdot A_{-}A_{+}^{*}\int_{1}^{\infty}\frac{dv}{v}\left[v^{2\nu} + \frac{1}{v^{2\nu}}\right]F(v) +\mathcal{O}\left(\lambda^2\right).
    \end{aligned}
\eeq
Here we see not only the logarithmic, but also the power-like divergence in the first terms. Such contributions to $n_{p}$ and $\varkappa_{p}$ also appear in EPP, but in that case they are exactly cancel each other in the correction to the Keldysh propagator (\ref{eq:Keldysh_at_large_times}). We see that in CPP for the modes with $p\eta > |\nu|$ this cancellation does not happen.

The second case that should be considered in CPP is for the modes with the momentum $p\eta < |\nu|$. In this case one cannot separate the corrections to $n_p$ and $\varkappa_p$ as in (\ref{eq:Keldysh_at_large_times}) and should consider the correction to the whole Keldysh propagator as in (\ref{eq:Keldysh_large_times_complementary_EPP}). Then the one-loop correction is expressed as
\beq\label{eq:one-loop_Keldysh_CPP}
    \Delta_{1-\text{loop}}G^{K}(\bm{p}|\eta,\eta) = -\frac{A_{-}^2}{(p\eta)^{2\nu}}\eta^{D-1}\cdot\lambda^2\log\left(\frac{\eta}{\eta_0}\right)\cdot 8A_{-}\text{Im}\left\{A_{+}\right\}\int_{1}^{\infty}\frac{dv}{v}v^{2\nu}\text{Im}\left\{F(v)\right\} + \mathcal{O}\left(\lambda^2\right),
\eeq
and the leading non-logarithmic corrections cancel each other, as it was the case in EPP.

Let us summarize and compare our observations for both in CPP and in EPP. In the case of EPP the integrals in (\ref{eq:1-loop_correction_general_formula_EPP}) are convergent, but have large contributions for such external momenta, $p$, that $p\eta \ll |\nu|$. For other momenta the corrections are just suppressed by the coupling constant, $\lambda^2$. 

At the same time, in CPP independently of the value of momentum $p$ such integrals as in (\ref{eq:1-loop_correction_general_formula_EPP}) are divergent at the lower bound of integration over internal times $\eta_{3,4}$, because the upper and lower limits of these integrals are exchanged in CPP, as compared to the case of EPP. One has to introduce the cutoff, $\eta_0$, at the lower bound of integration independently of the value of $p$. This is the position of the initial Cauchy surface. Furthermore, for such $p$ that $p\eta_0 < |\nu|$ there are large corrections to the Keldysh propagator. At the same time, for such $p$ that $p\eta_0 > |\nu|$ there are no large corrections. What is peculiar now, for such $p$ that $p\eta_0 < |\nu|$, but $p\eta > |\nu|$ ($\eta > \eta_0$) we have large power like (in $p\eta_0$) rather than logarithmic corrections, which are separately present in $n_p(\eta)$ and $\varkappa_p(\eta)$. At the same time for such $p$ that $p\eta < |\nu|$ we have only large logarithmic corrections, while power like corrections cancel each other in the whole expression for the corrected Keldysh propagator.  

Furthermore, it is straightforward to see that for the case\footnote{The case $p\eta > |\nu|$ should be considered separately.} $p\eta < |\nu|$ in CPP the diagrams of the type on the Fig.\ref{Fig:BubbleInsideBubble} will behave as $\lambda^4\log^2\left(\frac{\eta}{\eta_0}\right)$, as was shown in \cite{Akhmedov:2019cfd}. Therefore, in contrast with respect to the case of EPP, we must take into account the ``loops inside loops'' in the leading order in powers of the parameter $\lambda^2 \log\left(\frac{\eta}{\eta_0}\right)$. Hence, in the corresponding Dyson-Schwinger equation, which sums leading loop corrections, we should use the exact Keldysh propagator also inside internal lines. This will lead to a non-linear integral equation for the exact Keldysh propagator in CPP. As shown for principal series in \cite{Akhmedov:2013vka, Akhmedov:2012dn} the non-linear integral equations can lead to both singular solutions with the ``explosive'' behavior of the stress-energy tensor for a finite proper time and smooth solutions, depending on the initial conditions. The complication with the complementary series is inapplicability of the kinetic approximation, which works when the function $N(p\eta)$ (or $n_p(\eta)$ and $\varkappa_p(\eta)$) can be treated as slow function in comparison with the modes. We will return to the situation in CPP and in global dS for the fields from the complementary series in future work.

\section{Conclusion}\label{Sec:Conclusion}
In this paper we revised the secular effects in dS space-time, focusing on the complementary series. We derived the Dyson-Schwinger equation, which sums up the leading secular contributions in the case of light fields in EPP. We corrected the observations of \cite{Akhmedov:2017ooy} according to the observations that has been made in \cite{Akhmedov:2019cfd}. Namely, we excluded the diagrams of the type shown in Fig.\ref{Fig:BubbleInsideBubble}, as these diagrams are actually suppressed as compared to the bead or chaplet diagrams, and provide subleading contributions. 

Additionally, we have showed that the higher-point correlation functions should not be considered in the resulting system of Dyson-Schwinger equations, even for sufficiently small masses $\nu>\frac{D-1}{6}$, and can be taken to have its tree-level form. We then found the solution to this Dyson-Schwinger equation in one particular situation, when the initial values of quantum level populations and anomalous quantum averages are set at the initial physical momentum $P_0 = (p\eta)_{0}$. First, with this choice we preserve at least a part of the dS isometry, consisting of the rescalings $p\rightarrow \frac{1}{\alpha}p, \;\eta\rightarrow\alpha\eta $ (see also \cite{Starobinsky:1994bd, Tsamis:2005hd} for the related discussion). Second, it allows for a straightforward solution of the Dyson-Schwinger equation with a clear physical meaning. 

It is interesting to explore different forms of initial conditions in EPP. One option is to fix some initial time $\eta_0$, which can be considered as the time after which the self-interaction of the fields was adiabatically turned on, then define the initial values of $n_p(\eta_0)$ and may be even $\varkappa_p(\eta_0)$. However, taking $\eta_0$ to past infinity in this case is not possible, as it would lead to a divergent physical energy density. The second option is to use the momentum-dependent Bogolyubov coefficients when constructing modes (\ref{eq:field_decomposition}):
\beq\label{eq:Bogolyubov_coefficients}
    h_{\nu}(p\eta) \propto \alpha_{\bm{p}}H_{\nu}^{(1)}(p\eta) + \beta_{\bm{p}}H_{\nu}^{(1)}(p\eta).
\eeq
In this case we must suppose that the coefficient $\beta_{\bm{p}}$ approaches zero for large momenta rapidly enough to obtain propagators (for the Fock space ground state) with the proper Hadamard behavior. These options for the initial conditions violate the dS isometry at the tree-level. But their consideration still may provide further insights into the solutions of the Dyson-Schwinger equations for large IR contributions, depending on the initial state of the system at the beginning of its evolution.

In addition, we have considered loop corrections in CPP in Section \ref{Sec:Complementary_CPP}. While in the case of EPP the integrals in (\ref{eq:1-loop_correction_general_formula_EPP}) are convergent, but have large contributions for such external momenta, $p$, that $p\eta \ll |\nu|$, in CPP we encounter quite a different situation. 
In the latter case independently of the value of momentum $p$ such integrals as in (\ref{eq:1-loop_correction_general_formula_EPP}) are divergent at the lower bound of integration over internal times $\eta_{3,4}$, because the upper and lower limits of these integrals are exchanged in CPP, as compared to the case of EPP. One has to introduce the cutoff, $\eta_0$, at the lower bound of integration independently of the value of $p$. This is the position of the initial Cauchy surface. Furthermore, for such $p$ that $p\eta_0 < |\nu|$ there are large corrections to the Keldysh propagator. At the same time, for such $p$ that $p\eta_0 > |\nu|$ there are no large corrections. What is peculiar now, for such $p$ that $p\eta_0 < |\nu|$, but $p\eta > |\nu|$ ($\eta > \eta_0$) we have large power like (in $p\eta_0$) rather than logarithmic corrections, which are separately present in $n_p(\eta)$ and $\varkappa_p(\eta)$. At the same time for such $p$ that $p\eta < |\nu|$ we have only large logarithmic corrections, while power like corrections cancel each other in the complete expression for the corrected Keldysh propagator.  We will derive the Dyson-Schwinger equation, which sums these corrections up, in our future work and also consider the situation in global dS. 

Finally, in Appendix \ref{App:Principal_EPP} we demonstrated that for the principal series, unlike the complementary series, there is a static solution for the exact $n_{p}(\eta)$ and $\varkappa_{p}(\eta)$ when the scalar field has sufficiently large mass.

\section*{Acknowledgments}
AET would like to acknowledge discussions for many years on related issues with F.Popov and A.Polyakov.
The work of D.Sadekov was supported by the Foundation for the Advancement of Theoretical Physics and Mathematics ``BASIS''. This work was partially supported by the Ministry of Science and Higher Education of the Russian Federation (agreement no. 075–15–2022–287).

\appendix

\section*{A brief comment on the principal series in the expanding Poincare patch}\label{App:Principal_EPP}
Secular effects in EPP for the principal series were investigated in \cite{Krotov:2010ma, Akhmedov:2013vka}. In this appendix we just reconsider the derivation of the Dyson-Schwinger equation, making sure that the ``bubble inside bubble'' diagrams from Fig.\ref{Fig:BubbleInsideBubble} are not included as the corrections of subleading order. 

In the field quantization (\ref{eq:field_decomposition}) for principal series ($m > (D-1)/2$) the modes are as follows:
\beq\label{eq:modes_principal_series}
        f_{\bm{p}}(\eta) = \eta^{\frac{D-1}{2}}h_{i\nu}(p\eta), \;\;h_{i\nu}(p\eta) = H_{i\nu}^{(1)}(p\eta)\frac{\sqrt{\pi}}{2}e^{-\frac{\pi}{2}\nu}, \;\;\nu=\sqrt{m^2-\frac{D-1}{2}}.
\eeq
In this case the behavior of the Hankel function at small arguments are oscillatory rather than power-like:
\beq\label{eq:asymptotics_principal_series}
\begin{aligned}
    h_{i\nu}(p\eta)\simeq A_{+}(p\eta)^{i\nu}+A_{-}(p\eta)^{-i\nu},
    \\
    A_{+} = \frac{2^{-i\nu} e^{-\frac{\pi}{2}\nu}\sqrt{\pi}\left(1+\coth\left(\pi\nu\right)\right)}{2\Gamma\left(1+i\nu\right)}, \;\;
    A_{-} = -\frac{i2^{i\nu}}{2\sqrt{pi}}e^{-\frac{\pi}{2}\nu}\Gamma\left(i\nu\right).
\end{aligned}
\eeq
Then at one-loop order the corrections to the level populations and to the anomalous quantum averages are \cite{Krotov:2010ma, Akhmedov:2013vka}:
\beq\label{eq:n_and_kappa_principal_series}
\begin{aligned}
    n_{p}(\eta) \simeq -\Gamma\cdot e^{-2\pi\nu}\lambda^2\log(p\eta),\;
    \text{Re}\left\{\varkappa_{p}(\eta)\right\}\simeq \frac{\Gamma}{2}\cdot e^{-\pi\nu} \lambda^2\log(p\eta),
    \\
    \text{Im}\left\{\varkappa_{p}(\eta)\right\}\simeq C\cdot e^{-\pi\nu} \lambda^2\log(p\eta),
    \\
    \Gamma = \frac{S_{D-2}}{(2\pi)^{D-1}}\frac{1}{2\nu}\left|\int_{0}^{\infty}dx x^{\frac{D-3}{2}-i\nu}h_{i\nu}^{2}(x)\right|^{2},
    \\
    C = \frac{2}{\nu}\int_{1}^{\infty}\frac{dv}{v}\cos\left(\nu\log(v)\right)\int\frac{d^{D-1}\bm{\ell}}{(2\pi)^{D-1}}\text{Im}\left\{h_{i\nu}^2(\ell n)h_{i\nu}^{*2}\left(\frac{\ell}{v}\right)\right\},
\end{aligned}
\eeq
where $S_{D-2}$ is the volume of the unit $(D-2)$-dimensional sphere.
Please note here the presence of the extra suppressing factor $e^{-\pi \nu}$, which is present on top of the coupling constant $\lambda^2$. Furthermore, every higher loop brings extra power of this factor. Then for large masses there is perhaps no even necessity to sum up the loop corrections, because it demands too many e-foldings to make the loop corrections comparable to the tree-level contributions to the correlation functions.

Nevertheless let us consider the Dyson-Schwinger equation, which resums the leading corrections, because for sufficiently small $\nu$  the factor $e^{-\pi \nu}$ is essentially absent. Then in the kinetic approximation we can write the Dyson-Schwinger equation for the leading logarithms (\ref{eq:n_and_kappa_principal_series}) \cite{Akhmedov:2013vka}:
\beq\label{eq:DS-equation_Principal_series_EPP}
    \begin{aligned}
    G^{K}(\bm{p}|\eta,\eta) = G^{K}_{0}(\bm{p}|\eta,\eta)+ \lambda^2\int\frac{d^{D-1}\bm{k}}{(2\pi)^{D-1}}\frac{d\eta_{3}}{\eta_{3}^{D}}\frac{d\eta_{4}}{\eta_{4}^{D}}
    \bigg[-\frac{1}{2}\theta(\eta_3-\eta)\theta(\eta_4-\eta)\times&
    \\
    \times \rho\big(\bm{p}\big|\eta,\eta_3\big)\left(G^{K^2}_{0}\big(\bm{k}-\bm{p}\big|\eta_3,\eta_4\big) - \frac{1}{4}\rho^2\big(\bm{k}-\bm{p}\big|\eta_3,\eta_4\big)\right)\rho\big(\bm{p}\big|\eta_4,\eta\big) + &
    \\
    + \theta(\eta_4-\eta_3)\theta(\eta_3-\eta)\rho\big(\bm{p}\big|\eta,\eta_3\big)\rho\big(\bm{k}-\bm{p}\big|\eta_3,\eta_4\big) G^K_{0}\big(\bm{k}-\bm{p}\big|\eta_3,\eta_4\big)G^K\big(\bm{p}\big|\eta_4,\eta\big)
    + &
    \\
    +\theta(\eta_3-\eta_4)\theta(\eta_4-\eta)G^K\big(\bm{p}\big|\eta,\eta_3\big)\rho\big(\bm{k}-\bm{p}\big|\eta_3,\eta_4\big) G^K_{0}\big(\bm{k}-\bm{p}\big|\eta_3,\eta_4\big)\rho\big(\bm{p}\big|\eta_4,\eta\big)
    \bigg]&.
\end{aligned}
\eeq
In order to find an approximate solution to this equation we use the ansatz (\ref{eq:Keldysh_at_large_times}) for the exact Keldysh propagator and for simplicity choose the Bunch-Davies vacuum as the initial state. Then we have the following equations for $n_p(\eta)\equiv n(p\eta)$ and $\varkappa_p(\eta)\equiv \varkappa(p\eta)$:
\beq\label{eq:DS-equation_n_Principal_series_EPP}
    \begin{aligned}
        n(p\eta) \simeq -\Gamma \cdot e^{-2\pi\nu} \lambda^2\log(p\eta)\; +\; 
        \\
        +\lambda^2\int_{\eta}^{\infty}\frac{d\eta_4}{\eta_4}\int_{\eta_4}^{\infty}\frac{d\eta_3}{\eta_3}\left(\eta_3\eta_4\right)^{\frac{D-1}{2}}
        \text{Im}\bigg\{h_{i\nu}^{*}(p\eta_3)h_{i\nu}(p\eta_4)n\left(p\sqrt{\eta\eta_3}\right) + h_{i\nu}(p\eta_3)h_{i\nu}(p\eta_4)\varkappa\left(p\sqrt{\eta\eta_3}\right) \bigg\}\cdot
        \\
        \cdot\int\frac{d^{D-1}\bm{k}}{(2\pi)^{D-1}}\text{Im}\bigg\{h_{i\nu}(k\eta_3)h_{i\nu}\left(|\bm{k}-\bm{p}|\eta_3\right)h_{i\nu}^{*}(k\eta_4)h_{i\nu}^{*}\left(|\bm{k}-\bm{p}|\eta_4\right)  \bigg\} + \mathcal{O}\left(\lambda^2\right),
    \end{aligned}
\eeq
\beq\label{eq:DS-equation_kappa_Principal_series_EPP}
    \begin{aligned}
        \varkappa(p\eta) \simeq  \left[\frac{\Gamma}{2} + iC \right]\cdot e^{-\pi\nu}\lambda^2\log(p\eta)\;+\;
        \\
        +i\lambda^2\int_{\eta}^{\infty}\frac{d\eta_4}{\eta_4}\int_{\eta_4}^{\infty}\frac{d\eta_3}{\eta_3}\left(\eta_3\eta_4\right)^{\frac{D-1}{2}}
        \bigg[h_{i\nu}^{*}(p\eta_3)h_{i\nu}^{*}(p\eta_4)n\left(p\sqrt{\eta\eta_3}\right) + h_{i\nu}(p\eta_3)h_{i\nu}^{*}(p\eta_4)\varkappa\left(p\sqrt{\eta\eta_3}\right) \bigg]\cdot
        \\
        \cdot\int\frac{d^{D-1}\bm{k}}{(2\pi)^{D-1}}\text{Im}\bigg\{h_{i\nu}(k\eta_3)h_{i\nu}\left(|\bm{k}-\bm{p}|\eta_3\right)h_{i\nu}^{*}(k\eta_4)h_{i\nu}^{*}\left(|\bm{k}-\bm{p}|\eta_4\right)  \bigg\} + \mathcal{O}\left(\lambda^2\right).
    \end{aligned}
\eeq
The equations (\ref{eq:DS-equation_n_Principal_series_EPP})-(\ref{eq:DS-equation_kappa_Principal_series_EPP}), in contrast with the case of complementary series discussed in Section \ref{subSec:Resummation_complementary_EPP}, have static solution $n(p\eta)\simeq n_f,\;\varkappa(p\eta)\simeq \varkappa_f$ with:
\beq
    \begin{aligned}
        n_f \; \simeq \; \frac{\Gamma^2+3C^2}{\Gamma^2+C^2}\cdot e^{-2\pi\nu},
        \\
        \varkappa_f \;\simeq \;2\;\frac{\Gamma^2+2C^2 - i\Gamma C}{\Gamma^2+C^2}\cdot e^{-\pi\nu}.
    \end{aligned}
\eeq
This solution is meaningful if $n_f\ll 1$ and $\varkappa_f \ll 1$, which is certainly true under the condition $\nu \gtrsim \frac{2}{\pi}\log\left(\frac{1}{\lambda}\right)$. Therefore, in the case of sufficiently large masses of a scalar field with Bunch-Davies vacuum as the initial state, the system only slightly restructures the initial state and passes into some energy density on top of the out-vacuum as a result of a long evolution in the expanding background.

\bibliography{literature}

\begin{thebibliography}{10}

\bibitem{Starobinsky:1994bd}
Alexei~A. Starobinsky and Junichi Yokoyama.
\newblock {Equilibrium state of a selfinteracting scalar field in the De Sitter background}.
\newblock {\em Phys. Rev. D}, 50:6357--6368, 1994.

\bibitem{Tsamis:2005hd}
N.~C. Tsamis and R.~P. Woodard.
\newblock {Stochastic quantum gravitational inflation}.
\newblock {\em Nucl. Phys. B}, 724:295--328, 2005.

\bibitem{Marolf:2010zp}
Donald Marolf and Ian~A. Morrison.
\newblock {The IR stability of de Sitter: Loop corrections to scalar propagators}.
\newblock {\em Phys. Rev. D}, 82:105032, 2010.

\bibitem{Marolf:2010nz}
Donald Marolf and Ian~A. Morrison.
\newblock {The IR stability of de Sitter QFT: results at all orders}.
\newblock {\em Phys. Rev. D}, 84:044040, 2011.

\bibitem{Higuchi:2010xt}
Atsushi Higuchi, Donald Marolf, and Ian~A. Morrison.
\newblock {On the Equivalence between Euclidean and In-In Formalisms in de Sitter QFT}.
\newblock {\em Phys. Rev. D}, 83:084029, 2011.

\bibitem{Moreau:2018lmz}
G.~Moreau and J.~Serreau.
\newblock {Stability of de Sitter spacetime against infrared quantum scalar field fluctuations}.
\newblock {\em Phys. Rev. Lett.}, 122(1):011302, 2019.

\bibitem{Guilleux:2015pma}
Maxime Guilleux and Julien Serreau.
\newblock {Quantum scalar fields in de Sitter space from the nonperturbative renormalization group}.
\newblock {\em Phys. Rev. D}, 92(8):084010, 2015.

\bibitem{Hollands:2010pr}
Stefan Hollands.
\newblock {Correlators, Feynman diagrams, and quantum no-hair in deSitter spacetime}.
\newblock {\em Commun. Math. Phys.}, 319:1--68, 2013.

\bibitem{Gorbenko:2019rza}
Victor Gorbenko and Leonardo Senatore.
\newblock {$\lambda \phi^4$ in dS}.
\newblock 10 2019.

\bibitem{Akhmedov:2009vs}
Emil~T. Akhmedov and Philipp Burda.
\newblock {A Simple way to take into account back reaction on pair creation}.
\newblock {\em Phys. Lett. B}, 687:267--270, 2010.

\bibitem{Krotov:2010ma}
Dmitry Krotov and Alexander~M. Polyakov.
\newblock {Infrared Sensitivity of Unstable Vacua}.
\newblock {\em Nucl. Phys. B}, 849:410--432, 2011.

\bibitem{Akhmedov:2017ooy}
E.~T. Akhmedov, U.~Moschella, K.~E. Pavlenko, and F.~K. Popov.
\newblock {Infrared dynamics of massive scalars from the complementary series in de Sitter space}.
\newblock {\em Phys. Rev. D}, 96(2):025002, 2017.

\bibitem{Akhmedov:2019cfd}
E.~T. Akhmedov, U.~Moschella, and F.~K. Popov.
\newblock {Characters of different secular effects in various patches of de Sitter space}.
\newblock {\em Phys. Rev. D}, 99(8):086009, 2019.

\bibitem{Starobinsky:1980te}
Alexei~A. Starobinsky.
\newblock {A New Type of Isotropic Cosmological Models Without Singularity}.
\newblock {\em Phys. Lett. B}, 91:99--102, 1980.

\bibitem{Starobinsky:1982ee}
Alexei~A. Starobinsky.
\newblock {Dynamics of Phase Transition in the New Inflationary Universe Scenario and Generation of Perturbations}.
\newblock {\em Phys. Lett. B}, 117:175--178, 1982.

\bibitem{Linde:1981mu}
Andrei~D. Linde.
\newblock {A New Inflationary Universe Scenario: A Possible Solution of the Horizon, Flatness, Homogeneity, Isotropy and Primordial Monopole Problems}.
\newblock {\em Phys. Lett. B}, 108:389--393, 1982.

\bibitem{Linde:1983gd}
Andrei~D. Linde.
\newblock {Chaotic Inflation}.
\newblock {\em Phys. Lett. B}, 129:177--181, 1983.

\bibitem{Guth:1980zm}
Alan~H. Guth.
\newblock {The Inflationary Universe: A Possible Solution to the Horizon and Flatness Problems}.
\newblock {\em Phys. Rev. D}, 23:347--356, 1981.

\bibitem{Guth:1982ec}
Alan~H. Guth and S.~Y. Pi.
\newblock {Fluctuations in the New Inflationary Universe}.
\newblock {\em Phys. Rev. Lett.}, 49:1110--1113, 1982.

\bibitem{Albrecht:1982wi}
Andreas Albrecht and Paul~J. Steinhardt.
\newblock {Cosmology for Grand Unified Theories with Radiatively Induced Symmetry Breaking}.
\newblock {\em Phys. Rev. Lett.}, 48:1220--1223, 1982.

\bibitem{Polyakov:2007mm}
A.~M. Polyakov.
\newblock {De Sitter space and eternity}.
\newblock {\em Nucl. Phys. B}, 797:199--217, 2008.

\bibitem{Akhmedov:2008pu}
Emil~T. Akhmedov and P.~V. Buividovich.
\newblock {Interacting Field Theories in de Sitter Space are Non-Unitary}.
\newblock {\em Phys. Rev. D}, 78:104005, 2008.

\bibitem{Polyakov:2009nq}
A.~M. Polyakov.
\newblock {Decay of Vacuum Energy}.
\newblock {\em Nucl. Phys. B}, 834:316--329, 2010.

\bibitem{Akhmedov:2009be}
Emil~T. Akhmedov, P.~V. Buividovich, and Douglas~A. Singleton.
\newblock {De Sitter space and perpetuum mobile}.
\newblock {\em Phys. Atom. Nucl.}, 75:525--529, 2012.

\bibitem{Polyakov:2012uc}
A.~M. Polyakov.
\newblock {Infrared instability of the de Sitter space}.
\newblock 9 2012.

\bibitem{Akhmedov:2013vka}
E.~T. Akhmedov.
\newblock {Lecture notes on interacting quantum fields in de Sitter space}.
\newblock {\em Int. J. Mod. Phys. D}, 23:1430001, 2014.

\bibitem{Akhmedov:2021rhq}
E.~T. Akhmedov.
\newblock {Curved space equilibration versus flat space thermalization: A short review}.
\newblock {\em Mod. Phys. Lett. A}, 36(20):2130020, 2021.

\bibitem{Akhmedov:2013xka}
E.~T. Akhmedov, F.~K. Popov, and V.~M. Slepukhin.
\newblock {Infrared dynamics of the massive \ensuremath{\phi}4 theory on de Sitter space}.
\newblock {\em Phys. Rev. D}, 88:024021, 2013.

\bibitem{bunch1978quantum}
Timothy~S Bunch and Paul~CW Davies.
\newblock Quantum field theory in de sitter space: renormalization by point-splitting.
\newblock {\em Proceedings of the Royal Society of London. A. Mathematical and Physical Sciences}, 360(1700):117--134, 1978.

\bibitem{kamenev2023field}
Alex Kamenev.
\newblock {\em Field theory of non-equilibrium systems}.
\newblock Cambridge University Press, 2023.

\bibitem{Akhmedov:2012dn}
E.~T. Akhmedov.
\newblock {Physical meaning and consequences of the loop infrared divergences in global de Sitter space}.
\newblock {\em Phys. Rev. D}, 87:044049, 2013.

\bibitem{polyakov2012infraredinstabilitysitterspace}
A.~M. Polyakov.
\newblock Infrared instability of the de sitter space, 2012.

\bibitem{Higuchi_2011}
Atsushi Higuchi, Donald Marolf, and Ian~A. Morrison.
\newblock Equivalence between euclidean and in-in formalisms in de sitter qft.
\newblock {\em Physical Review D}, 83(8), April 2011.

\bibitem{hollands2010correlatorsfeynmandiagramsquantum}
Stefan Hollands.
\newblock Correlators, feynman diagrams, and quantum no-hair in desitter spacetime, 2010.

\bibitem{Akhmedov:2022uug}
E.~T. Akhmedov, I.~V. Kochergin, and M.~N. Milovanova.
\newblock {Isometry invariance of exact correlation functions in various charts of Minkowski and de Sitter spaces}.
\newblock {\em Phys. Rev. D}, 107(10):105015, 2023.

\bibitem{Akhmedov:2024ice}
E.~T. Akhmedov and K.~A. Kazarnovskii.
\newblock {On two approaches to quantization in strong background fields}.
\newblock 10 2024.

\bibitem{Gautier:2013aoa}
Florian Gautier and Julien Serreau.
\newblock {Infrared dynamics in de Sitter space from Schwinger-Dyson equations}.
\newblock {\em Phys. Lett. B}, 727:541--547, 2013.

\bibitem{Guilleux:2016oqv}
Maxime Guilleux and Julien Serreau.
\newblock {Nonperturbative renormalization group for scalar fields in de Sitter space: beyond the local potential approximation}.
\newblock {\em Phys. Rev. D}, 95(4):045003, 2017.

\bibitem{Serreau:2013psa}
Julien Serreau and Renaud Parentani.
\newblock {Nonperturbative resummation of de Sitter infrared logarithms in the large-N limit}.
\newblock {\em Phys. Rev. D}, 87:085012, 2013.

\end{thebibliography}
\bibliographystyle{unsrt}
\end{document}